\documentstyle[psfig,12pt]{article}
\def\reference{\parskip 0pt\par\noindent\hangindent 0.5 truecm}
\newcommand{\HI}{H$\,${\sc i}}

\newcommand{\kms}{km\,s$^{-1}$}
\begin{document}
%
%
\title{The Violent Interstellar Medium of Nearby Dwarf Galaxies}
%


\author{Fabian Walter
} 

\date{}
\maketitle
{\center
Radioastronomisches Institut, Bonn, Germany\\E--mail: walter@astro.uni--bonn.de\\[3mm]
}

%
\begin{abstract}
High resolution \HI\ observations of nearby dwarf galaxies (most of
which are situated in the M\,81 group at a distance of about 3.2\,Mpc)
reveal that their neutral interstellar medium (ISM) is dominated by
hole--like features most of which are expanding. A comparison of the
physical properties of these holes with the ones found in more massive
spiral galaxies (such as M\,31 and M\,33) shows that they tend to
reach much larger sizes in dwarf galaxies. This can be understood in
terms of the galaxy's gravitational potential. The origin of these
features is still a matter of debate. In general, young star forming
regions (OB--associations) are held responsible for their
formation. This picture, however, is not without its critics and other
mechanism such as the infall of high velocity clouds, turbulent
motions or even gamma ray bursters have been recently proposed. Here I
will present one example of a supergiant shell in IC\,2574 which
corroborates the picture that OB associations are indeed creating
these structures. This particular supergiant shell is currently the
most promising case to study the effects of the combined effects of
stellar winds and supernova--explosions which shape the neutral
interstellar medium of (dwarf) galaxies.
\end{abstract}

{\bf Keywords:}
galaxies: individual (IC 2574, Holmberg II, DDO 47, NGC 3077), ISM: kinematic and dynamics, ISM: structure, radio lines: ISM, X--rays: ISM
\bigskip

%
%

\section{On Holes and Shells in Galaxies}
Since the discovery by Heiles (1979, 1984) of huge, shell--like
structures in the \HI\ distribution of our Galaxy similar such
structures have been found in several nearby galaxies, such as in
M\,31 (Brinks \& Bajaja 1986), M\,33 (Deul \& den Hartog 1990),
Holmberg\,II (Puche et al.\ 1992) and the Small Magellanic Cloud
(Staveley--Smith et al. 1997). These discoveries have been made
possible due to the advent of powerful aperture synthesis radio
telescopes such as the Very Large Array (VLA), the Westerbork
Synthesis Radio Telescope (WSRT) and the Australia Telescope Compact
Array (ATCA). One of the most stunning examples of what is possible
these days is without doubt the latest ATCA image of the LMC (Kim et
al.\ 1998b).

All these observations indicate that the interstellar medium (ISM) of
medium to late--type galaxies is dominated by features, which have
variously been described as shells, rings, holes, loops, bubbles or
cavities (the 'cosmic bubble bath' (Brand \& Zealey 1975) or 'violent
interstellar medium'). Until recently, most authors concentrated on
large spiral systems. However, there are several advantages in using
dwarf galaxies to study the structure of the ISM: dwarfs are slow
rotators, generally show solid body rotation, and lack density
waves. This implies that once features like shells have formed, they
won't be deformed by galactic shear and therefore tend to be long
lived. Moreover, the overall gravitational potential of a dwarf is
much smaller than that of a normal spiral, hence expanding structures
can more easily reach large sizes.

Figure~1 (top left) shows an \HI--map of Holmberg\,II, a member of the
M\,81 group (at a distance of about 3.2\,Mpc) by Puche et al.\
(1992). This was the first high resolution, high sensitivity \HI\
image of a nearby dwarf galaxy ever obtained and shows that the
presence of \HI\ holes clearly dominates its overall appearance (in
total, 51 \HI\ holes were catalogued by Puche et al.). This result
prompted several questions such as: Is Holmberg\,II a special case? Do
other dwarf galaxies show the same structures in their ISM?

Figure~1 (top right) presents a VLA \HI\ surface brightness map of the
dwarf galaxy IC\,2574, also a member of the M\,81 group (taken from
Walter \& Brinks 1999). As can be seen from this map, the ISM of
IC\,2574 is clearly shaped by \HI\ holes and shells -- a total of 48
were catalogued by Walter \& Brinks. Figure~1 (bottom left) shows the
distribution of neutral hydrogen in DDO\,47 (Walter \& Brinks 1998b)
which is a dwarf galaxy situated at an assumed distance of
4\,Mpc. Again, many holes and shells are visible in this galaxy -- in
fact, DDO\,47 resembles Holmberg\,II quite a bit. So far, about 20
\HI\ holes have been catalogued in this galaxy, most of which are
expanding (Walter \& Brinks, in prep.).

So do all nearby dwarf galaxies show similar structures?  To answer
this question, studies of other dwarf galaxies are needed. Here I will
briefly discuss NGC\,3077, another member of the M\,81 group. The
distribution of the neutral hydrogen in this galaxy is presented in
Fig.~1 (bottom right). Like all the other maps presented here, this
map was obtained with the VLA after combining observations in three
different configurations (B--, C-- and D--array). An analysis of the
data rather surprisingly showed that no expanding hole--like features
are visible in this particular galaxy (Walter et al., in prep.).  Note
however, that NGC\,3077 is a special case since it is strongly
interacting with its high--mass neighbors M\,81 and M\,82 --- it may
therefore very well be that structures like holes and shells are
destroyed prematurely due to the strong tidal forces.

\begin{figure}[t] 
    \vspace{-0.5cm}
\caption{
{\em Top left:} \HI\ surface brightness map of Holmberg\,II (Puche et
al.\ 1992). The greyscale is a linear representation of the \HI\
surface brightness. In total, 51 \HI\ holes were catalogued by Puche
et al.  {\em top right:} \HI\ surface brightness map of IC\,2574
(Walter \& Brinks 1999). The greyscale is a linear representation of
the \HI\ surface brightness. We find a total of 48 \HI\ holes in this
galaxy, most of which are expanding.  {\em bottom left:} \HI\ surface
brightness map of DDO\,47. The greyscale is a linear representation of
the \HI\ surface brightness. A companion galaxy (not visible) has been
discovered somewhat south--east of the main galaxy (see Walter \&
Brinks, 1998b).  {\em bottom right:} \HI\ surface brightness map of
NGC\,3077 (Walter et al., in prep.). The greyscale is a linear
representation of the \HI\ surface brightness. The map has not been
corrected for primary beam attenuation.  The two prominent tidal arms
connect this galaxy to M\,81 and M\,82 (see Yun, Ho \& Lo 1994).}
\end{figure}

\section{Holes in Dwarfs vs.\ Holes in Spirals}

Having convinced ourselves that the most distinctive features of the
ISM in dwarf galaxies are \HI\ holes and shells, at least provided
they are not participating in strong interactions, the question is now
how their physical properties compare to the holes found in more
massive, spiral galaxies. A detailed comparison has been performed by
Brinks \& Walter (1998) and only the main results are highlighted
here.  In what follows the observed \HI\ hole properties of M\,31
(Brinks \& Bajaja 1986), an example of a massive spiral galaxy similar
to our own; M\,33 (Deul \& den Hartog 1990), a less massive spiral;
IC\,2574 (Walter \& Brinks 1999) and Ho\,II (Puche et al.\ 1992), two
dwarf galaxies in the M\,81 group of galaxies (note that Ho\,II is
four times less massive than IC\,2574) will be compared. In other
words, the sequence M\,31 -- M\,33 -- IC\,2574 -- Ho\,II spans a large
range of different Hubble types from massive spirals to low--mass
dwarfs.

Despite the fact that a similar study, and at considerably higher
sensitivity, is now available for the SMC (Staveley--Smith et al.\
1997), we have decided not to include their results, the main reason
being that the linear scales (or spatial frequencies) sampled by the
ATCA only just overlap with those observed in the galaxies listed
above. The linear resolution, at 28 pc, is almost four times higher
than, for example, the VLA maps of IC\,2574. At the other end of the
spectrum, because of the lack of short spacing information, structures
larger than a few hundred parsec will have been missed. Moreover, the
SMC is a very disturbed system, being torn apart by tidal forces due
to interactions with the LMC and the Galaxy. An additional argument
for leaving out the SMC is that Staveley--Smith et al.\ used a
different approach in searching and identifying the holes which makes
a direct comparison difficult.

Other objects with catalogued \HI\ holes, such as DDO\,47 (Walter \&
Brinks 1998b) and IC\,10 (Wilcots \& Miller 1998) have so few holes
that a statistical analysis is not warranted.  Limiting the comparison
to the four galaxies listed above has some further advantages. The
linear resolutions are very similar, as are the velocity resolutions
with which they have been observed (see Walter \& Brinks 1999 for
details). In addition, all four galaxies were examined in more or less
the same fashion, one of the authors (E.\ Brinks) having taken part in
the analysis of three of the four objects.  The results for the four
galaxies suffer partially from low statistics and incompleteness due
to personal bias and observational constraints (such as the
beamsize). However, these effects, to first order, affect a comparison
in a similar way and that it is valid to try to find global trends as
a function of Hubble type. In order to remove the human factor, it
would be interesting to apply an automated object recognition package
such as that developed by Thilker et al.\ (1998) to all galaxies with
sufficiently detailed observations.

Fig.~2 (left) shows an overlay of the relative size distribution of
the holes found in the four galaxies. In this plot the bins are on a
linear scale. Note that there is a clear sequence with Hubble type!
The size distribution for holes in M\,31 and M\,33 cuts off sharply
near 600 pc. In contrast, holes in IC\,2574 and Ho\,II reach sizes of
1200 to 1500 pc, respectively. The lack of holes with sizes smaller
than $\sim 100$\,pc is due to our resolution limit. As explained in
the previous section, holes are larger for ``later'' Hubble types
because these smaller galaxies have lower masses and hence a lower
mass surface density.  So, for the same amount of energy deposited, an
\HI\ shell can grow much larger, both because of a lower gravitational
potential and a lower ambient density. Because the \HI\ layer is much
thicker in dwarfs as well (Puche et al.\ 1992; Walter \& Brinks 1999,
b), shells take longer to break out of the disk.

From the observed hole properties such as the expansion velocities and
diameters, one can try to estimate the amount of energy which was
needed to produce the holes, based on numerical simulations. Here we
use the numerical model developed by Chevalier (1974) to calculate the
total mechanical energy needed to create the holes. Note that many
assumptions enter the calculations and that the results should only be
taken to be order of magnitude estimates. A plot of the results of
this analysis is shown in Fig.~2 (right). Gratifyingly, the energies
needed to produce the holes are the same for all galaxies -- this
suggests that, whatever the underlying physical mechanism for the
creation of the holes is, it seems to be the same in all galaxies, at
least to first order.

In addition to the objects listed in the introduction and the maps
shown here, several more dwarf galaxies have been observed. Walter \&
Brinks are working on data on Holmberg\,I and M\,81\,dwarf\,A. Van Dyk
et al.\ (1998) present data for Sextans\,A. Wilcots
and collaborators have data on three more galaxies, IC\,1613, IC\,10, and
NGC\,4449, the former object being completely dominated by \HI\ holes
and shells, much like IC\,2574 and Ho\,II. Hence, within the coming
year, a lot more material should thus become public, allowing for
better comparative studies to be performed.

\begin{figure}[t] 
\caption{{\em Left:} Comparison of the relative distribution, in percentage, of
the diameters of the \HI\ holes in IC\,2574, M\,31, M\,33 and Ho\,II.
{\em right:} Comparison of the relative distribution, in percentage, of
the energies required to produce the \HI\ holes in IC\,2574, M\,31, M\,33
and Ho\,II.}
\end{figure}

\section{So, what created the holes?}

Until not too long ago, the origin of these structures was generally
thought to lie in the combined effects of stellar winds and supernova
explosions produced by young stellar associations. For review articles
see Tenorio--Tagle \& Bodenheimer (1988), and van der Hulst (1996, and
references therein). However, several authors have pointed out that
this model is not without its flaws. One of the potential problems
with the hypothesis that the \HI\ holes are the result of an evolving
OB association in which the most massive stars, through their winds
and supernovae, create the observed supershells is the
following. Using reasonable assumptions one still expects a
substantial population of A and F main sequence stars to be
present. However, searches by Radice et al.\ (1995) and Rhode et al.\
(1997) in galaxies like Ho\,II have not led to the expected result.  A
possible alternative explanation has been proposed by Efremov,
Elmegreen \& Hodge (1998) who suggest that Gamma Ray Bursters, which
recently have been conclusively associated with objects at
cosmological distances, might provide the required energy, and occur
frequently enough, to explain the observed \HI\ supershells.

Another objection to the standard model is that in the case of the
largest observed shells, the energy requirements seem to exceed the
output of stellar winds and supernovae.  To explain those structures
an alternative mechanism was proposed: the infall of gas clouds.
Tenorio--Tagle et al.\ (1987) present a numerical simulation and van
der Hulst \& Sancisi (1988) provide what is probably the best
observational evidence for infall, the case of one of the largest
holes in M\,101.

One should also consider the possibility that we are tricked by nature
and that the holes that we see to be expanding are actually the result
of turbulent motions. A search for their powering sources would then
be completely futile. To investigate this possibility I examined
turbulence cubes calculated by Mac--Low et al.\ (1998b) in the very same
fashion as was done in our search for \HI\ holes in galaxies (Walter \&
Brinks 1999). The result was that only a few percent of the smaller
holes may be due to turbulence. However, in the case of the larger
holes ($>100$\,pc), turbulence is not able to produce coherent
features which seem to be expanding. We therefore feel confident that
the structures we observe are indeed due to expanding \HI\ shells.

Obviously, to investigate the sources which created the holes, a
multi--wavelength approach is needed. In this approach, 21\,cm
observations are needed for the identification of the holes as well as
for the determination of their kinematics. Optical observations are
indispensable to check the stellar distribution and populations within
the shells. Narrow band H$\alpha$ observations are important to trace
the regions of current star formation. Quite often,
H$\alpha$--emission is found to be located close to or on the rim of
the holes, as defined by the \HI--observations. Finally, X--ray
observations are important to check whether the cavities of the
\HI\ holes are filled by hot X--ray emitting gas or not. A hot--gas
interior is one of the main predictions of theories which state that
the holes are created by young OB--associations (see also the
discussion below).

So far, only a few shells have been found where such an approach is
possible.  Examples are the supergiant shell LMC\,4 (Bomans, Dennerl
\& K\"urster 1994), the superbubbles N\,44 (Kim et al.\ 1998a) and
N\,11 (Mac~Low et al.\ 1998a), all three situated in the LMC, the
supergiant shell SGS\,2 in NGC\,4449 (Bomans, Chu \& Hopp 1997) and
the possible supershell near Holmberg\,IX (Miller 1995). In the
following section, the detection of a supergiant shell in the nearby
dwarf galaxy IC\,2574 is presented which is probably the most
prominent supergiant shell known to date. This region has proved to be
an ideal laboratory to study the physical nature of supergiant shells
in general and is expected to provide conclusive proof as regards to
the source lying at its origin.

\section{The case of the Supergiant shell in IC\,2574}

The supergiant shell in IC\,2574 was first seen in high resolution VLA
\HI\ observations (Walter \& Brinks 1999; see also Fig.\ 3, left
panel, which is a scaled--down version of Fig.\ 1 (top right).  The
shell has a linear size of about 1000\,pc $\times$ 500\,pc
($60''\times30''$) and is expanding at $\sim$25\,\kms. It is therefore
an ideal target to study expansion models since despite its size it
has not stalled yet (as most of the supergiant shells in the LMC
have). The elliptical shape of the \HI\ shell is indicated in Figs.~3
and 4. The kinematic age based on the observed size and expansion
velocity is estimated at 14 Myr.

Deep narrow--band H$\alpha$--imaging revealed that current star
formation (SF) regions within IC\,2574 are predominantly situated on
the rim of the \HI\ shell (Fig.~3, right panel, greyscale).  This suggests
that we are witnessing triggered star formation on the rim due to the
expansion of the \HI\--shell (see, e.g., Elmegreen 1994). Follow--up
radio continuum observations showed that these starforming regions are
the main source of the radio continuum emission (see the contours in
Fig.~4 for a map of the $\lambda$6\,cm emission).

Various theories on the creation and formation of supergiant shells
(SGSs) predict that the cavity within the shell should be filled with
hot gas (see, e.g., Cox \& Smith 1974, Weaver et al.\ 1977, Chu et
al.\ 1995).  A pointed ROSAT observation towards IC\,2574 (Walter et
al.\ 1998) revealed that the supergiant shell is indeed filled with
extended hot X--ray gas (see the contours in Fig.\,3).  This makes the
supergiant shell in IC\,2574 a truly unique region and suggests that
we have caught this SGS at an auspicious moment. Assuming a
Raymond--Smith (1977) plasma temperature of log($T\rm
[K])\,=\,6.8\pm0.3$ and an internal density of
$(0.03\,\pm\,0.01)$\,cm$^{-3}$ we derive an internal pressure of
$P\approx 4 \times10^5$\,K\,cm$^{-3}$.  This pressure is much higher
than the pressure of the ambient warm ionized medium ($P\approx
10^3-10^4$\,K\,cm$^{-3}$) suggesting that it is probably this hot gas
which is still driving the expansion of the shell (see, e.g., Weaver
et al.\ 1977).

\begin{figure}[t]
\caption[]{Left: IC\,2574 in the 21\,cm line of neutral hydrogen
(\HI). Right: Blowup of the supergiant shell in IC\,2574. The ellipses
plotted in both maps indicate the size of the expanding \HI\ shell
(linear size $\approx 1000\times 500$\,pc).  The greyscale is a
representation of the H$\alpha$ emission coming from the rim of this
shell. The contours represent the X--ray emission coming from the
inside of the shell as observed with the ROSAT PSPC camera (for
details see Walter et al.\ 1998). Coordinates are given for B1950.0.}
\end{figure}

\begin{figure}[t]
      \caption[]{The supergiant shell in IC\,2574, showing the same
      region as in Fig.~3 (right panel). The plotted ellipse again indicates
      the size of the expanding \HI\ shell. The greyscale is a
      representation of our deep R--band image showing the central
      stellar association. The superimposed contours represent
      $\lambda$6\,cm radio continuum emission.}
\end{figure}

We have just been granted observing time during Cycle~1 with the
Advanced X-ray Astrophysics Facility (AXAF) so we will soon be able to
derive the spatial extent and the temperature of the X--ray gas to a
much higher accuracy.  The AXAF observations will also allow us to
determine the contribution to the X--ray flux by point sources (e.g.,
X--ray binaries and supernovae).  Note that the X--ray source is
resolved in the ROSAT observations, indicating that at least a
significant fraction of the X--ray emission is extended.

From ground based R--band imaging, a giant stellar association is
readily visible within the IC\,2574--SGS (see Fig.~4, greyscale).  We
speculate that this stellar association is in fact responsible for the
formation and expansion of the shell as well as for the heating of the
X--ray gas. Unfortunately, the evidence is still largely
circumstantial.

The wealth of observations which is available for this supergiant
shell suggests that this central stellar association is the powering
source for the formation and expansion of the shell as well as for the
heating of the X--ray gas.  Based on our \HI\ observations and using
the models of Chevalier (1974), we derive that the energy required to
produce the shell must be of order $10^{53}$ ergs or the equivalent of
about 100 Type II SNe. This would mean that the least massive stars
that go off as SN are most probably still present in the central
stellar association since their lifetimes ($\sim$ 50 Myr) are somewhat
longer then the dynamical age of the hole ($\sim$ 14 Myr, as derived
from the \HI\ observations).

\section{Conclusions}

\HI\ observations of sufficient angular and velocity resolution as
well as sensitivity of dwarf irregular galaxies are now becoming
available in the literature. Because dwarfs have several advantages
over spiral galaxies (the absence of density waves, the absence of
differential rotation and hence shear) one can expect a lot of progress
to be made in our understanding of the structure of the ISM. In
addition, multi--wavelength studies of the most prominent shells will
reveal which physical processes are the cause of the \HI\ holes in
general. The prominent supershell in IC\,2574 is a nice example of what
can be done in that respect. In summary:

\begin{enumerate}
\item Dwarf galaxies show a stunning amount of detail in the form of
shells and holes in their neutral interstellar medium.  These features are
similar to those found in large systems like our Galaxy, M\,31, M\,33,
M\,101 and NGC\,6946. Often, the \HI\ shells completely dominate the
morphology of a dwarf galaxy, such as in the case of IC\,2574 or Ho\,II.

\item The shells can grow to large dimensions because of several
conditions which are favourable in dwarf galaxies. The volume density
in the plane is low, which facilitates expansion. In the direction
perpendicular to the disk, the gravitational pull is smaller than in a
massive spiral. Also, because of the thick \HI\ layer, shells are
easily contained and unlikely to break out into the halo. Lastly,
solid body rotation and a lack of spiral density waves prevent holes
from being rapidly destroyed.

\item A comparison of IC\,2574 with other galaxies spanning a range of
Hubble types and studied in similar detail so far (M\,31, M\,33 and
Ho\,II) shows that the size distribution of \HI\ holes found in a
galaxy is related to its Hubble type in the following way. The size of
the largest \HI\ shells is inversely proportional to the global
gravitational potential (and hence mass surface density). The energies
needed to create these structures, though, are found to be of the same
order for all types of galaxies. Hence, whatever physical mechanism lies at
their origin, it is not related to the host galaxy, at least to first
order.

\item In order to determine the source(s) which created the observed
structures a multi--wavelength approach is needed. So far, a few
theories have been invoked to explain the presence of the holes: 1)
young OB--associations which drive the expansion of the shell by
strong stellar winds and supernova explosions, 2) Gamma Ray Bursters
and 3), for the largest holes, the infall of high velocity clouds. The
scenario regarding the creation due to strong stellar winds and
subsequent supernova explosions still has some problems. Detailed
studies of single \HI\ holes are needed to distinguish between the
various theories.

\item The supergiant shell in IC\,2574 seems to be an ideal target to
shed light on this problem. This supergiant shell is clearly defined
in \HI\ observations and is surrounded by massive star formation. A
pointed ROSAT observation has revealed that the cavity enclosed by the
supergiant shell is filled with hot, X--ray emitting gas. A prominent
stellar association in the center of this SGS is thought to be the
powering source for the formation and expansion of the shell as well
as for the heating of the interior X--ray emitting gas. Future
investigations, especially regarding the central stellar association,
are needed to understand fully what created this fascinating region.
\end{enumerate} 

%
%





\section*{Acknowledgements}


I would like to thank my thesis supervisor, Elias Brinks, for
invaluable help. I would also like to thank Ulrich Klein, Evan
Skillman and Neb Duric for fruitful discussions. I appreciate the help
of J\"urgen Kerp regarding the X--ray observations. I
acknowledge the Deutsche Forschungsgemeinschaft (German Science
Foundation, DFG) for the award of a fellowship in the Graduate School
'The Magellanic Clouds and Other Dwarf Galaxies'.

\section*{References}






\reference Brand, P.W.J.L., Zealey, W.J. 1975, A\&A, 38, 363

\reference Brinks, E., \& Bajaja, E. 1986, A\&A, 169, 14

\reference Brinks, E., \& Walter, F. 1998, in Proceedings of
	the Bonn/Bochum--Graduiertenkolleg Workshop "The Magellanic
	Clouds and Other Dwarf Galaxies", eds. T. Richtler and
	J.M. Braun, in press

\reference Bomans, D.J., Dennerl, K., \& K\"urster, M. 1994, A\&A, 283, L21

\reference Bomans, D.J., Chu, Y.--H., \& Hopp, U. 1997, AJ, 113, 1678

\reference Chevalier, R. A. 1974, ApJ, 188, 501

\reference Chu, Y.--H., Chang, H.--W., Su, Y.--L., \& Mac Low, M.--M. 1995,
ApJ, 450, 156

\reference Cox, D.P., \& Smith, B.W. 1974, ApJL, 189, L105

\reference Deul, E.R., \& den Hartog, R.H. 1990, A\&A, 229, 362

\reference Efremov, Yu.N., Elmegreen, B.G., \& Hodge, P.W. 1998, ApJL, accepted

\reference Elmegreen, B.G. 1994, ApJ, 427, 384

\reference Heiles, C. 1979, ApJ, 229, 533

\reference Heiles, C. 1984, ApJS, 55, 585

\reference Kim, S., Staveley--Smith, L., Dopita, M.A., Freeman, K.C.,
Sault, R.J., Kesteven, M.J., \& McConnell, D., 1998a, ApJ, 503, 674

\reference Kim, S., Chu, Y.--H., Staveley--Smith, L., \& Smith, R.C. 1998b, ApJ, 503, 729

\reference Mac Low, M.--M., Chang, T.H., Chu, Y.--H., \& Points, 
S.D. 1998a, ApJ, 493, 260

\reference Mac Low, M.--M., Klessen, R.S., Burkert, A.. \& Smith,
M.D. 1998b, Phys.\ Rev.\ Letters, 80, 2754

\reference Miller, B.W. 1995, ApJL, 446, L75

\reference Puche, D., Westpfahl, D., Brinks, E., \& Roy,
J.--R. 1992, AJ, 103, 1841 

\reference Radice, L.A., Salzer, J.J., \& Westpfahl, D.J. 1995,
BAAS, 186, 49.08 

\reference Rhode, K.L., Salzer, J.J., \& Westpfahl, D.J. 1997,
BAAS, 191, 81.09 

\reference Raymond, J.C., Smith, B.W. 1977, ApJ Supp., 35, 419

\reference Staveley--Smith, L., Sault, R.J., Hatzidimitrou, D.,
Kesteven, M.J., \& McConnell, D. 1997, MNRAS, 289, 225 

\reference Tenorio--Tagle, G., Franco, J., Bodenheimer, P., \&
R\'o\.zyczka, M. 1987, A\&A, 179, 219 

\reference Tenorio--Tagle, G., \& Bodenheimer, P. 1988, ARA\&A,
26, 145 

\reference Thilker, D.A., Braun, R., \& Walterbos, R.A.M. 1998, A\&A, 332, 429

\reference van der Hulst, J. M. 1996, in ASP
Conf. Ser. Vol. 106: The Minnesota Lectures on Extragalactic Neutral
Hydrogen, ed. E. D. Skillman (ASP:San Francisco), 47

\reference van der Hulst, J. M., \& Sancisi, R. 1988, AJ, 95, 1354

\reference Van Dyk, S.D., Puche, D., Wong, T. 1998, AJ, accepted

\reference Walter, F. \& Brinks, E., 1999, AJ, accepted

\reference Walter, F. \& Brinks, E., 1998, in Proceedings of IAU
  Colloquium 171 ``The Low Surface Brightness Universe'', Cardiff, UK

\reference Walter, F., Kerp, J., Duric, N., Brinks, E., \& Klein, U. 1998,
ApJL, 502, L143

\reference Weaver, R., McCray, R., Castor, J., Shapiro, P., \& Moore, 
R. 1977, ApJ, 218, 377

\reference Wilcots, E.M., \& Miller, B. 1998, AJ, 116, 2363

\reference Yun, M.S., Ho, P.T.P., \& Lo, K.Y. 1994, Nature, 372, 530

\end{document}